\documentclass[]{spie}

\usepackage{amsmath,amsfonts,amssymb}
\usepackage{graphicx}
\usepackage[colorlinks=true, allcolors=blue]{hyperref}
\usepackage{braket}
\usepackage{booktabs}
\usepackage{tabularx}
\usepackage{array}
\usepackage{longtable}
\usepackage{ragged2e}
\usepackage{caption}
\usepackage{xcolor}
\usepackage{parskip}

\title{\vspace{0.15in}
Eliminating photon transport in long-baseline optical\\
interferometry using quantum memories}

\author[a]{Yousef K. Chahine}
\author[b]{Chaohan Cui}
\author[b]{William DeRocco}
\author[b,c]{Daniel Gottesman}
\author[b,c]{Saikat Guha}
\author[c,d]{Emil T. Khabiboulline}
\author[b,c]{Zhenning Liu}
\author[e]{Brittany McClinton\textsuperscript{*}}
\author[e]{Jayadev Rajagopal}
\author[e]{Fredrik Rantakyro}
\author[b]{J. Gabriel Richardson}
\author[e]{Stephen Ridgway}
\author[b]{Aqil Sajjad}
\author[b]{Joohyung Song}
\affil[a]{NASA Glenn Research Center, Cleveland, Ohio, USA}
\affil[b]{University of Maryland, College Park, Maryland, USA}
\affil[c]{Joint Center for Quantum Information and Computer Science, NIST/University of Maryland, College Park, Maryland, USA}
\affil[d]{Joint Quantum Institute, NIST/University of Maryland College Park, Maryland, USA}
\affil[e]{NSF NOIRLab, Tucson, Arizona, USA}

\authorinfo{
\textsuperscript{*}Corresponding author: Brittany McClinton, \texttt{brittany.mcclinton@noirlab.edu}
}

\begin{document} 

\maketitle

\begin{abstract}

Current generation optical interferometers support telescope separations (baselines) up to $\sim$ 300 meters. Investigations of some astrophysical targets (e.g., stars, galactic nuclei, exoplanets) would benefit from considerably longer, even multi-kilometer, baselines. 

The realization of very long-baseline optical interferometry faces a number of challenges: transmission of light over large distances for beam recombination; an extreme precision continuously variable optical delay line; and low overall system efficiencies of $\sim 0.01$. With today’s natural extension of existing technologies, this requires extreme low-loss optical fibers, a massive, complex opto-mechanical infrastructure, and ever-larger apertures. However, even assuming state-of-the-art, classical methods make optical long-baseline interferometry impractical beyond a few kilometers.

All of these limitations can be addressed and potentially resolved with new technologies from quantum physics. Recent proposals (Gottesman, 2012; Khabiboulline, 2019) have suggested using entanglement, together with quantum memory (QM), to efficiently measure photon coherences across distant telescope sites. QM also promises to eliminate optical delay lines, substituted by controlled timing in memory loading, correlation, and tunable phase operations. These new avenues offer solutions for presently unachievable requirements of optical interferometry at longer baselines. Finally, capture of wavefront information in QM eliminates the need for the very large numbers of optical surfaces in contemporary facilities, with the potential to deliver a high fraction of the celestial photon flux to memory, and possible improvements in net efficiency.
 
In this paper, we present the fundamental operating mechanisms of optical interferometry using QM and entanglement. We show how these remove the optical delay line bottleneck. QM is not without its own set of challenges, some of which include very small bandwidths as well as limitations in storage time. We highlight keystone areas of technology that require further development and are essential to realizing these opportunities, as well as ongoing work to overcome these challenges.

\end{abstract}

\newpage

\section{Introduction}

\subsection{Classical context and limitations}
Current generation optical/infrared interferometers support telescope separations up to hundreds of meters, with baselines of order $\sim 300$~m representing the scale of present facilities. Many astrophysical targets, including stellar surfaces, galactic nuclei, and exoplanet systems, would benefit from substantially longer baselines, potentially extending to multi-kilometer scales. Such scaling of classical long-baseline optical interferometry faces barriers in the present day:
\begin{enumerate}
\item Direct-detection interferometry requires coherent transport of the astronomical optical field to a common beam combiner. Over multi-kilometer baselines, a number of factors progressively reduce throughput and sensitivity, including diffraction, finite optical apertures, multiple reflections, and (for guided transport) fiber attenuation.

\item Beam recombination requires the propagating optical fields to overlap within their coherence envelope while maintaining sub-wavelength differential path stability. It is increasingly difficult to satisfy this requirement over longer baselines due to the need for variable optical delay, chromatic dispersion compensation, polarization control, and precision metrology.

\item Free-space beam transport requires an uninterrupted line of sight between telescopes and the beam combiner. On baselines of many kilometers, the need for long evacuated beam paths or buried infrastructure over terrain make such systems increasingly difficult and expensive to construct and maintain.
\end{enumerate}
Straightforward extrapolation of existing infrastructure to multi-kilometer optical baselines would require low-loss beam transport, large and complex opto-mechanical systems, and increasingly large collecting apertures, making such facilities impractical with present classical methods.

\subsection{Quantum approaches}

Quantum-enhanced long-baseline interferometry approaches have been proposed that employ entanglement, a unique property of quantum objects. There are proposals which operate with or without some form of quantum memory, a device that can store quantum states and is used to access inter-site coherence without optical transport. A number of proposals and early demonstrations have established the basic feasibility of these approaches.

In one class of quantum-enhanced approaches, first proposed by Gottesman et al.~\cite{Gottesman2012LongBaseline}, entanglement is used to transmit the information from incoming photons over long baselines. Transport of light still occurs, but it is the entangled light, which can be replaced, controlled, and purified using quantum repeaters. The stellar light that contains the signal is not transported. Furthermore, delay lines are still required just as in a classical system, although again acting on the entangled light.
    
In a second class of approaches, first proposed by Khabiboulline et al.~\cite{khabiboulline_quantum-assisted_2019, khabiboulline_optical_2019}, the quantum state of the stellar light is transduced onto quantum memories. An ingredient which they propose is to do this transduction in a compressive fashion over many temporal modes during which one stellar photon is expected to arrive. With the assistance of entanglement, we measure which temporal mode the stellar photon arrived in, and hence which memories contain the relevant information about the photon. 

Compared to the first class of approaches, optical transport is only required for producing the banks of entangled qubits that are employed in the protocol and not for transporting the stellar light. Furthermore, delay lines are not necessary since the memories themselves can function as an equivalent time delay. An important consequence of using memories is that the amount of entanglement required decreases dramatically, making the scheme more practical, in that the load on quantum networks (the infrastructure for distributing entanglement) is reduced. Sophisticated quantum algorithms can be run on the quantum state of the stellar light stored in quantum memories by applying quantum computing and quantum networking operations on these quantum memories ~\cite{Padilla2024, Padilla2025}. Besides the Khabiboulline et al. proposal, there have been other approaches suggested or implemented for using quantum memory \cite{Bland-Hawthorn_2021,wang2026memory}. 

Both classes of approaches have been realized experimentally in simulated lab settings~\cite{Brown_2023, Stas_2026}.

\subsection{Scope of this paper}
This paper is intended as a conceptual and technical bridge between quantum-information protocols for optical interferometry and the practical requirements of astronomical long-baseline arrays. We focus on quantum-memory interferometry in which astronomical optical coherence is captured by quantum memory locally at each telescope and subsequently read out using entanglement-assisted operations. We describe the operating principles of the architecture in language familiar to the optical/infrared interferometry community, and identify how the usual interferometric requirements are transformed when propagating optical fields are replaced by phase-referenced quantum memories. 

The central question addressed in this paper is how the use of quantum memory changes timing and phase requirements, and what it leaves unchanged. Quantum memories can remove the need to physically transport astronomical photons to a central beam combiner, and can replace full-range optical delay lines with controlled memory loading, timing association, and phase operations. However, they do not eliminate the need to measure optical coherence, preserve phase information, compensate deterministic geometric delay, or manage atmospheric and instrumental phase fluctuations. We therefore recast timing, delay, and phase-stability requirements as operating budgets for a quantum-memory interferometer, distinguishing coarse synchronization and coherence-envelope overlap from optical phase fidelity and local reference stability.

This paper is structured as follows: 
\begin{itemize}
    \item Section \ref{sec:overview_narrative} provides an accessible overview for an astronomy audience of how a quantum-memory based long-baseline interferometer operates.
    \item Section \ref{sec:mathematical} develops timing and phase requirements for a quantum-memory based interferometer, providing results for efficiency of photon capture into memory and visibility based on phase and timing parameters. 
    \item  Section \ref{sec:keystone} describes keystone areas of technology development needed to realize a quantum-memory system in the near future for functional observations. 
    \item Section \ref{sec:conclusions} then wraps up the paper by providing conclusions and outlooks based on the results of the previous sections. 
\end{itemize}

\section{OPERATING PRINCIPLES} \label{sec:overview_narrative}

\subsection{Classical Interferometry as a Reference}
Classical optical/infrared interferometry measures spatial coherence by physically superposing the optical fields collected at separated apertures. After compensation of the geometric delay, the fields from telescopes A and B are brought to a common beam combiner, where the detected intensity contains an interference term proportional to the first-order mutual coherence \cite{Goodman_2005},
\begin{equation}
G^{(1)}_{AB}(t) = E_A^{(+)}(t) E_B^{(+)*}(t).
\end{equation}

For a spectrally filtered stellar field with coherence time ($t_c$), observable interference first requires temporal overlap of the arriving coherence envelopes, so that $|t_A-t_B| \lesssim t_c$. This is the familiar role of optical delay lines: they align the wave packets from different apertures so that the mutual coherence can be measured.

Envelope overlap, however, is not by itself sufficient for direct-detection beam combination. Because the transported optical fields are interfered at optical frequencies, residual optical-path fluctuations during the measurement interval enter as a carrier-phase error. If $\delta\tau(t)$ is the residual differential delay after compensation, the averaged fringe term is suppressed unless
\begin{equation}
\left\langle e^{i\omega_0 \delta\tau(t)} \right\rangle_{T_{\text{int}}} \neq 0
\end{equation}
remains appreciably nonzero over the detector averaging interval ($T_{\text{int}}$). Thus classical direct detection imposes two simultaneous requirements: overlap of the coherence envelope and sub-wavelength stability of the relative optical path during the interval over which the fringe is coherently measured.

Quantum-memory interferometry seeks to recover the same complex visibility, but changes where coherence is stored and where interference is performed. What changes is that the propagating optical fields need not be transported to a central beam combiner; instead, the coherence may be mapped locally into phase-referenced quantum memory states and read out later through entanglement-assisted operations. As discussed below, this shifts the relevant stability condition from preservation of direct optical-carrier interference between transported fields to control of the residual phase in a locally referenced memory frame, together with preservation of the coherence-envelope overlap needed to identify the corresponding temporal modes. The coherence envelope overlap is achieved in this case by correctly aligning temporal modes at each site using entangled pairs. We will expound later on an analogy between quantum memory based interferometry and heterodyne interferometry. In heterodyne interferometry, the averaging interval is sensitive to the residual baseband phase error $\delta\phi_{res}(t)$ in the rotating frame rather than to an unconverted optical-frequency beat between transported fields. In quantum memory interferometry, the averaging interval is now formed from locally referenced rotating-frame memory modes rather than through baseband voltages.

\subsection{Quantum Memory Capture of a Wavefront} \label{sec:qmem_load}

  In a quantum-memory interferometer, the optical mode collected at each telescope interacts locally with a memory qubit. The goal is to transfer the coherent superposition of `photon at telescope A' and `photon at telescope B' into a corresponding coherent superposition of memory states at the two locations.

This type of transfer can be described using standard quantum optics operations, such as: (a) an entangling operation between the optical mode and the memory; (b) a measurement of the optical mode; and (c) depending on the measurement outcome of the optical mode, a possible correction operation on the quantum memory to complete the state transfer~\cite{NielsenChuang2010}. In one physical embodiment of quantum memory, the entangling operation is realized with a reflective atom-cavity interface of the Duan-Kimble type~\cite{DuanKimble2004}. In such a system, the incoming photon reflects off of a cavity containing a memory qubit, acquiring a $\pi$ phase if the memory is in $\ket{\downarrow}$, while remaining unaltered if the  memory is in $\ket{\uparrow}$. Operationally, this provides an entangling gate between the optical qubit and the local memory qubit.

For a single baseline, an incident stellar photon in a selected temporal-spectral mode may be written schematically as a superposition of one photon at telescope A and one photon at telescope B, with a relative phase difference, $
\frac{1}{\sqrt{2}}(\ket{1_A 0_B} + e^{i \phi} \ket{0_A 1_B})$, where $1/0$ indicate the presence or absence of a photon, the subscripts indicate the telescope sites $A/B$, and $\phi$ is the astronomical visibility phase for the baseline between telescopes $A$ and $B$. After a successful loading procedure, the photon state is transferred into the memories as
\begin{align} 
\ket{\Psi_m} = \frac{1}{\sqrt{2}}
\left(
\ket{\overline{1}_A \overline{0}_B}
+
e^{i\phi}
\ket{\overline{0}_A \overline{1}_B}
\right)
\label{eq:memory_phase_transfer}
\end{align}

where the barred states denote state-transferred memory excitations rather than propagating photons. 

For simplicity, the expression above describes one temporal mode and one baseline. In the full protocol proposed by Khabiboulline et al.~\cite{khabiboulline_quantum-assisted_2019,khabiboulline_optical_2019}, the same idea is applied to many possible arrival-time bins. The state of the stellar light over $M$ temporal modes is stored in $\log_2(M+1)$ memories at each site according to a logarithmic compression scheme. Entanglement-assisted parity checks are then used to identify which memories contain the photon based on the temporal mode in which the photon arrived. This is done without revealing which telescope received the photon, allowing the interferometric information to be preserved and read out later. This replaces long optical delay lines with memory loading, timing association, and quantum measurement. This process will be discussed in more detail in the context of timing and phase requirements in Section~\ref{sec:mathematical}.

The above mentioned heralding measurement during the state transfer process involves measuring whether the optical modes at the two locations are in $\frac{1}{\sqrt{2}}(\ket{1_A 0_B} +\ket{0_A 1_B})$ or in $\frac{1}{\sqrt{2}}(\ket{1_A 0_B} -\ket{0_A 1_B})$. This can be conducted by carrying out what is called a single-rail x-basis measurement on each optical mode, and comparing the outcome from each telescope site~\cite{khabiboulline_optical_2019, khabiboulline_quantum-assisted_2019}: the same outcome means $+$, a different outcome means $-$. This x-basis measurement can be conducted probabilistically in present linear-optical implementations, and different schemes trade success probability against experimental complexity~\cite{Kok2010,Sajjad2026-single-rail}. A different approach that distinguishes between $\frac{1}{\sqrt{2}}(\ket{1_A 0_B} \pm \ket{0_A 1_B})$ involves supplying the same coherent state to both sites~\cite{khabiboulline_quantum-assisted_2019}. A recent solid-state quantum-network experiment has demonstrated the relevant memory-assisted nonlocal interferometry sequence in the laboratory~\cite{Stas_2026}. 

\subsection{Visibility Reconstruction}

First-order coherence is extracted using quantum operations on the stored memories. The configuration of the memories, dependent on the arrival time of the photon, is first determined in a way that does not destroy the coherence using entanglement-assisted parity checks to perform a non-local time-of-arrival measurement. Afterward, the coherence can be read out. As described previously, in classical optical interferometry the complex visibility is commonly reconstructed either through controlled optical path modulation or through simultaneous quadrature measurements. Different optical phase offsets probe different projections of the mutual coherence function, allowing both visibility amplitude and phase to be determined. Quantum-memory interferometry approaches this differently, although it is equivalent mathematically. Rather than mechanically varying the delay between propagating optical fields, the optical coherence is interrogated through different quantum measurement operations. Different measurement bases access different projections, or quadratures, of the stored coherence. Phase shifts may be applied during the memory interrogation or readout process, analogous to varying the phase of a local oscillator in homodyne detection.

Individual photon events produce only binary probabilistic outcomes. The complex visibility emerges statistically through ensemble averaging over many repeated measurements. In this sense, the quantum measurements play a role analogous to phase stepping or quadrature detection in classical interferometry. The interferometric information is reconstructed from accumulated statistical correlations, similar to a directly observed optical fringe. For multiple apertures, coherent quantum measurements over the full array provide advantages, such as improved signal-to-noise (S/N), over pairwise classical visibility measurements.

The process of visibility reconstruction for a single baseline will be described in more detail later in the context of timing and phase stability in section \ref{sec:two_site}.

\subsection{Multimode Interferometry to Attain Quantum Resolution Limit for Arbitrary Imaging Problems}

A possible alternative to the standard pairwise combination of telescopes for estimating complex visibilities is to perform a joint measurement on the light collected by more than two telescopes. Quantum estimation theory provides powerful tools for evaluating the performance of a given measurement scheme for a specified parameter-estimation or object-classification task. It also allows us to determine the maximum performance permitted by the laws of physics for a given scene, thereby guiding the search for measurements that are optimal for imaging, parameter estimation, or object classification. Relevant analytical tools include the classical and quantum Fisher information for parameter estimation, the classical and quantum Chernoff exponents for object classification~\cite{Helstrom1976}, and Bayesian adaptive methods such as Personick's minimum mean-squared error approach~\cite{Personick1971}. In general, the optimal measurement may require interfering signals from several telescopes jointly, rather than combining telescopes only pairwise to estimate the complex visibilities followed by a nonlinear digital inversion.

In the non-entanglement-based version, the light collected at all $n$ telescopes would be physically transported to a central location, where it would be mixed in a linear interferometer with $n$ inputs and $n$ outputs~\cite{Sajjad2024}. This amounts to measuring the telescope signals in a basis of their orthogonal linear combinations, which can be chosen using quantum-estimation-theoretic tools to optimize the performance for a given imaging or parameter-estimation task. Such a measurement, involving arbitrary linear combinations of signals from several telescopes rather than only pairwise combinations, would be practically very challenging to implement using physical beam combination with existing technologies---even more so than conventional pair-wise optical interferometry. However, an entanglement-based approach may eventually provide a way to overcome this barrier. Once the photonic state collected by multiple telescopes is loaded onto quantum memories at the different locations, performing a joint measurement on those memories in a suitably chosen collective basis becomes a distributed quantum computing task. With sufficient advances in quantum technology, this could allow the implementation of the optimal collective measurements suggested by quantum estimation theory. The conceptual framework for such arbitrary measurements on multiple telescopes has been described~\cite{Padilla2024, Padilla2025}, and could eventually allow interferometric imaging systems to surpass the performance obtainable from pairwise telescope combinations alone.

A further advantage of using quantum memories may be derived from performing collective measurements on the joint state of several incoming photons stored in quantum memories, rather than measuring each photon individually and independently. Quantum estimation theory shows that such collective measurements can outperform schemes where we only measure the individual photons independently, even if the individual photon measurements are chosen optimally~\cite{Deshler-unpublished}.

\section{Phase and Timing Stability} \label{sec:mathematical}

In this section, we consider timing and phase stability requirements for a quantum memory based interferometer. We present results for the probability that a photon of coherence time $t_c$ is captured by a bank of memories exposed in a sequence of time bins of width $t_w$ equal to the inverse bandwidth of the memory. This probability that the photon is loaded contributes not only to the overall inefficiency in the interferometric array, but also the maximum visibility as a function of the time bin synchronization error $\tau$ relative to the memory exposure bin width $t_{w}$ for various coherence time ratios $t_c/t_w$. These results are presented without rigorous derivation, which will be included in a future peer-reviewed publication, as the intention of this paper is to describe quantum-memory based long-baseline interferometry requirements for an astronomy audience without assuming a rigorous background in quantum information. We also discuss how various system-level phase artifacts are handled in a quantum memory based system, and make direct analogy of some operating requirements to heterodyne interferometry. Finally, we use these findings to present a table of requirements for a quantum memory based interferometer. 

\subsection{Setup}

We suppose that the light incident on the telescope array consists of a thermal field in a well-defined plane-wave spatial mode restricted to a single polarization, but which consists of an equiprobable mixture over all spectral-temporal modes.  Regardless of the spectral content of the source, this approximation is assumed to be valid in the vicinity of a single spectral channel with optical bandwidth on the order of 1-10 GHz in which an interferometric measurement will be performed.  We assume the broadband stellar light is filtered using some type of resonator to separate spectral channels before delivery to the memories. As a result, the filtered wave packet is modeled as an exponentially-decaying single-photon wave packet with coherence time $t_c$.  

Let $t_c\ll T\ll t_c/\epsilon$, where $\epsilon$ is the photon occupancy number, denote some duration much larger than the coherence time but less than the mean time between photon arrivals within the spectral channel.  We can approximate the state of the incident field within the time window $T$, including our uncertainty in the photon arrival time, as a mixed state of the form
\begin{equation}
\hat\rho = (1-p_1)|0\rangle\langle 0| + \frac{p_1}{T}\int_{0}^{T} |\psi_t\rangle\langle\psi_t| dt
\end{equation}
where $\ket{0}$ represents no photon, $\ket{\psi_t}$ represents the filtered wavepacket with leading edge at time $t$, and $p_1\ll 1$ is the probability that a photon arrives within the time window $0\leq t\leq T$.

We suppose that quantum memories are exposed to the incident field for some finite duration $t_w$ during which a photon may arrive. The process of coupling incident light into a quantum memory can be understood through direct analogy to the process of coupling light into an optical fiber. For example, just as the fiber provides unoccupied modes for an incident photon to excite, an absorptive atom-cavity quantum memory system can be understood as providing unoccupied modes for the excitation (a reflective quantum memory system operates in a somewhat different manner). In this analogy, the propagation axis of the fiber is replaced by the time axis of the atom-cavity system, and the spatial-frequency content of the incident free-space field is replaced by its temporal-frequency content. 
The atom-cavity system is modeled as a two-level system which can only unitarily couple light from a single time-frequency mode of the incident field, just as a single-mode fiber can only unitarily couple light from a single spatial mode. Similar to the manner in which coupling into single-mode fiber can be modeled using a pupil-plane overlap integral with the fiber mode back-propagated to the aperture of a focusing system, we can model coupling to the memory via an overlap integral with the time-frequency mode of the gated atom-cavity system back-propagated to the optical domain.  

For the following results, we model the back-propagated time-frequency mode of the gated atom-cavity system as a square-wave packet matched to the gate interval in the optical domain.  This corresponds to a frequency mode described by a sinc function of width $\sim$1/$t_w$.  Although this is not a particularly realistic resonant mode for an atom-cavity system, it is a convenient model with the property that any other eigenmode would yield strictly smaller overlap with an incident field gated within the exposure window, analogous to coupling light from a hard circular aperture using a fiber engineered to guide an Airy pattern mode.\cite{GRISSANCHEZ2016} Finally, the back-propagated memory mode is mapped onto the physical atom-cavity system using a unitary operation that imprints a relative phase $\theta_{\textup{LO}}$ from a local oscillator used to load the memory.

\subsection{Photon capture efficiency}

The probability of a photon capture during a sequence of memory exposures of total duration $T$ is
\begin{equation}
\boxed{P(\textup{capture}) = p_1\frac{2t_c}{t_w}\left(1 - \frac{t_c}{t_w}[1 - e^{-t_w/t_c}]\right)}
\end{equation}
which also yields the capture efficiency $\eta_c = P(\textup{capture})/p_1$.  This expression leads us to define 3 distinct operating regimes based on the ratio of the filtered photon coherence time and the memory eigenmode coherence time:

\begin{itemize}
    \item \textbf{Matched memory bandwidth ($t_w = t_c$):} If the memory exposure window is chosen to match the filtered stellar photon state, the capture efficiency is given by $\eta_c = 2/e$.  This inefficiency arises from mode-mismatch between the temporal mode of the filtered photon and memory.  We note that this is fundamentally caused by the uncertainty in the stochastic arrival time location of the temporal mode envelope of the particular stellar photon draw and thus cannot be avoided by matching the mode shape (e.g. modeling a square wave packet of width $t_c$ matching the memory eigenmode yields a similar capture efficiency $\eta_c=2/3$).
    \item \textbf{Underfilled memory bandwidth ($t_w \ll t_c$):} The memory exposure window is short, meaning the memory acceptance bandwidth is broad relative to the filtered stellar photon state. Here, the capture efficiency approaches unity ($\eta_c\to 1$) at the cost of requiring \emph{more memories}, leading to a less efficient use of memory bandwidth.
    \item \textbf{Overfilled memory bandwidth ($t_c \ll t_w$):} The memory acts as a narrow filter relative to the stellar photon state, yielding a correspondingly lower capture efficiency due to spectral mismatch, limiting to $\eta_c\sim 2t_c/t_w$ as $t_c/t_w\to 0$.  Here, utilization of available memory bandwidth is maximized, but photons outside the memory bandwidth may be discarded.
\end{itemize}

\subsection{Interferometric visibility with time bin synchronization error} \label{sec:two_site}

If we were to readout the state of the memories locally at each site after loading, we would obtain a vanishingly small averaged visibility as we would be averaging over the time bins which do not overlap with the arrival of a photon.  Instead, the Khabiboulline et al~\cite{khabiboulline_optical_2019} protocol first projects the state onto a definite pair of time bins using a non-local time-of-arrival measurement.  This can be achieved by correlating the $j$-th time bins at each site using pre-shared entangled quantum states, colloquially called ``Bell pairs".  

Bell pairs must be distributed to each site by the end of the total exposure time $T$, and must be stored locally at each site in a memory platform which can realize a controlled phase flip (CZ) gate from the memories used to store the stellar photon.  After distributing the Bell pairs, a series of CZ gates from each memory qubit onto the associated Bell pair entangles the state of the memories with the Bell pairs.
After this entangling operation, the Bell pairs are measured in order to recover the bin in which the Bell pair phase was flipped.  This is achieved by measuring each pair locally in the diagonal basis, and later identifying the time bins where opposite measurement results were obtained.

\begin{figure}[h]
\centering
\includegraphics[scale=0.7]{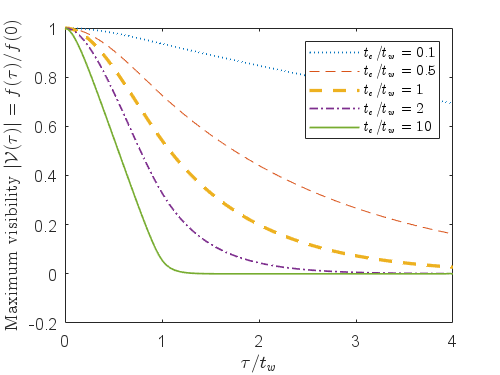}
\caption{Maximum visibility of two-site interference with time bin synchronization error $\tau$ relative to the memory exposure bin width $t_w$ for various coherence time ratios $t_c/t_w$.  Based on analysis of a spectrally-filtered weak thermal field loaded into a sequence of temporally-gated quantum memories.}
\label{fig:vlbicoherence}
\end{figure}

After the CZ-gate from a memory at site $A/B$ to an associated Bell pair at site $A/B$ is enacted, that memory can be immediately read out. The local measurements of the Bell pairs are later reconciled to recover a time bin $j$. If the detections with the local oscillators failed to yield successful click patterns when loading the $j$-th memory time bin at each site, then we record no result for the total exposure time $T$.  If no photon arrives during the total exposure time $T$, the Bell pair measurements will indicate that there was no arrival since none of the Bell pairs flipped and no result is recorded.  Assuming that the capture efficiency is the same at both sites, the density matrix between sites $A$ and $B$ reduces to the form
\begin{equation}
\hat\rho_j = \begin{pmatrix} 1/2 & \mathcal V/2 \\ \mathcal V^*/2 & 1/2 \end{pmatrix}
\end{equation}
where $\mathcal V$ is the interferometric complex visibility given by
\begin{align}\label{eq:visibility}
\boxed{\mathcal V(\tau) = e^{i\omega_0\Delta t}e^{i(\theta_{\textup{LO},B}-\theta_{\textup{LO},A})}\frac{f(\tau)}{f(0)}}
\end{align}
where $\tau=|t_{A,j}-t_{B,j}-\Delta t|$ is the time bin synchronization error, $\Delta t$ is the propagation delay between sites for the incident plane wave mode, $\omega_0$ is the center frequency of the spectral channel coupled to memory, and 
\begin{align}
f(\tau) = \frac{2t_c}{t_w}\left(e^{-t_w/t_c}\cosh\Big(\frac{\tau}{t_c}\Big) - e^{-\tau/t_c} + \frac{[t_w-\tau]_+}{t_c}-\sinh\Big(\frac{[t_w-\tau]_+}{t_c}\Big)\right)
\label{eq:coherencemagnitude}
\end{align}

where $[x]_+$ denotes the positive part of $x$ (i.e., $[x]_+=x$ if $x>0$ and vanishes otherwise). Note that the visibility degrades if the oscillator phase difference $\theta_{\textup{LO},B}-\theta_{\textup{LO},A}$ is not stabilized during the entire course of the measurement integration time $T_{\textup{int}}\gg T$, or if path loss and capture efficiency between sites $A$ and $B$ are not equalized.  Figure \ref{fig:vlbicoherence} shows the maximum visibility $|\mathcal V|$ based on the time-bin synchronization error between sites assuming perfectly stable phase-locked local oscillators and equal path-loss loading into the quantum memories.

We make two final remarks about the nature of the timing and phase stability requirements for this scheme:
\begin{itemize}
\item The phase stability requirements are decoupled from the coherence envelope overlap timing requirement described by Figure \ref{fig:vlbicoherence}, in contrast to a classical interferometer where the same optical path length determines both.  In the scheme above, the coherence envelope overlap is determined by timing the switch that gates the memories and choosing which time bins to entangle with Bell pairs, while the phase stability is independently determined by the stability of the phase-locked \textup{LO}s used to load the memories.
\item The Bell pairs carry no intrinsic information relevant to the interferometric phase.  In particular, entanglement distribution can be performed using dual-rail photonic qubits~\cite{knill_scheme_2001} over non-phase-stable links, provided they can be loaded into a platform which supports the required CZ-gate with the stellar-photon memories.
\end{itemize}

\subsection{Handling Other Phase Artifacts} \label{sec:conceptual_clarification}

In any optical interferometer, the measured phase difference between two telescope signals contains several distinct contributions,
\begin{equation}
\phi_{\rm total}
=
\phi_{\rm astro}
+
\phi_{\rm offset}
+
\phi_{\rm geom}
+
\phi_{\rm nuisance},
\end{equation}
where, 
\begin{itemize}
    \item $\phi_{\rm astro}$ is the desired astronomical visibility phase determined by the source structure and the spatial-frequency sampling of the array
    \item $\phi_{\rm offset}$ collects a variety of terms that are effectively constant, corresponding to a small position change of the source on the sky that is not relevant to usual imaging measurements
    \item $\phi_{\rm geom}$ is the deterministic phase associated with geometric propagation delay and Earth rotation. For a large optical baseline, it can change rapidly. However, it is also very predictable, except for slow drifts. In classical systems, this delay is compensated actively with a moving delay line which is controlled at the sub-wavelength level.  The required maximum delay is similar to the telescope separation. For hypothetical multikilometer systems, this becomes implausible for optical, mechanical and topographic reasons. In the quantum approach, the predictable, fast varying component can be tracked using the LO phase given the close relation of the memory loading to a classical heterodyne system as summarized in section \ref{sec:heterodyne}.
    \item $\phi_{\rm nuisance}$ term represents unwanted and mostly unpredictable phase fluctuations introduced by instrument and geophysical drift, and atmospheric turbulence.
\end{itemize}

With quantum memory support, shutter timing and quantum gates on the quantum memory replace the delay line. The geometric delay between telescope sites results in an offset of encoding times in memories, which can be synchronized digitally (e.g., by shifting the time bins). The ``memory shutter" must merely be timed at each telescope so as to capture all or most of the coherence packet. The measured coherence amplitude is weakly dependent on the timing of the shutter window, as described in the previous sub-section. The recovered phase is largely insensitive to the precise placement of the shutter window provided that both sites capture the coherence packet. Known phase artifacts, such as $\phi_{\text{geom}}$, can be corrected in the quantum memory using quantum gates. One class of gates which can be applied to the quantum state is the phase gate $U_{\theta}$, which realizes the operation
\begin{align}
    U_{\theta} \ket{0} = \ket{0} \\
    U_{\theta} \ket{1} = e^{i \theta} \ket{1}
\end{align}
For some state $\ket{\phi} = \frac{1}{\sqrt{2}}(\ket{0} + e^{-i \theta} \ket{1})$, applying such an operation gives
\begin{align}
    U_{\theta} \ket{\phi} = \frac{1}{\sqrt{2}}(\ket{0} + \ket{1}) \,.
\end{align}
It is easy to see that if the quantum state were $\frac{1}{\sqrt{2}}(\ket{0} + e^{i(\phi_{\text{astro}} + \phi_{\text{geom}})}\ket{1})$, a phase gate where $\theta = -\phi_{\text{geom}}$ would correct the state to the desired $\frac{1}{\sqrt{2}}(\ket{0} + e^{i\phi_{\text{astro}}}\ket{1})$. Returning to the nuisance term $\phi_{\text{nuisance}}$, the fastest and most problematic nuisance term is usually atmospheric piston. The aperture-averaged piston relevant to optical path difference (OPD) control evolves on the crossing time of the wind $V$ past the telescope aperture $D$, roughly as $\tau_{\rm piston}\sim 0.3D/V$. In both classical and quantum, this fluctuating OPD component presents a challenge. Quantum interferometry has no fundamental advantage over classical methods in this regard. Closure phases and other similar properties that are immune to atmospheric errors are used in classical interferometry as constraints in the inversion from the Fourier to image domain. In an accompanying paper at this conference, Gorshkov et al. begin to consider equivalent quantum measurements\cite{Gorshkov2026_SPIE}.

Classical systems determine the fringe phase by measuring the interference signal at multiple known OPD offsets, either simultaneously or through controlled modulation. Measured phase drift is used to generate feedback to the delay line. In order to achieve the required sub-wavelength control, servo bandwidths up to a few kHz may be required. With this control established, it is possible to vary the delay systematically and determine the actual OPD. Then it is usual to drive the delay to zero or near-zero OPD (ZPD). This supports calibrating the coherence amplitude which, away from ZPD, is depressed by averaging over the spectral bandwidth. A practical quantum implementation will likewise require repeated measurements at different effective OPD offsets in order to reconstruct the complex visibility. In the quantum approach, the equivalent of delay shifts can be implemented electronically during synchronous read out of the memories. Quantum-based techniques still require that the phase evolution be measured or calibrated on timescales shorter than $\tau_{\rm piston}$, but this requirement is considerably more relaxed than the high-bandwidth servo control needed for classical fringe tracking.  

If equipped with a (very short) optical delay line, quantum-based techniques could follow the same strategy as classical, using precise laboratory OPD offsets, but with substantially lower requirements on photon rate. However, the class of quantum memories considered in this paper inherently functions with narrow optical bandwidth per spectral channel.  This will greatly reduce the dependence of coherence amplitude on OPD. The visibility phase remains sensitive to OPD, but the coherence envelope becomes much broader, reducing amplitude loss from residual delay errors. Further, a multispectral quantum system could use the  phase shift with wavelength of multiple channels to continuously measure OPD for use in calibrating amplitudes. In such a configuration, precise real-time control of OPD may no longer be required. 

In summary, quantum memories remove the requirement for kilometer-scale optical delay lines and relax the timing precision needed to capture the incoming wavefront. They do not remove the need to determine the coherence phase and calibrate the amplitude within intervals of phase stability. In quantum-based techniques the challenge shifts from active optical-path control during measurement to synchronized clocks with accurate and stable timing and phase referencing during memory acquisition, storage and readout. Quantum memory may support other benefits to interferometry, For instance: the large geometric delay between telescopes may be learned with fewer photons using algorithms on the memories~\cite{Liu2026TimeDelays}; possible increases in efficiency (classical is notoriously inefficient); quantum-optimum strategies for measuring closure phase efficiently\cite{Gorshkov2026_SPIE}; and selectivity, which is to accept photons from the intended target and to suppress others. These topics are for a future study.  

\subsection{Relation to Heterodyne Interferometry}\label{sec:heterodyne}

A helpful way to understand the operating requirements of a quantum memory based interferometer is to compare the operating mechanism to the well-known classical technique of heterodyne interferometry. In heterodyne interferometry, the signal radiation is mixed, at each of the N sites, with a Local Oscillator (LO) that provides a stable phase reference. The resulting intermediate frequency (IF) carrier is detected and subsequently correlated between all pairs of apertures. The detected IF current, I(t), carries the relative phase between the signal, $E_{s}$ and the LO, $E_{LO}$ at the difference frequency.
\begin{equation}
I(t) \propto E_s E_{LO} \cos\left[ (\omega_s - \omega_{LO})t + (\phi_s - \phi_{LO}) \right]
\label{eq:IF_current}
\end{equation}
Hence, for the successful measurement of the correlation, the phase of the LO (the reference phase distributed over the array) has to be stable over timescales on the order of the inverse optical bandwidth. Heterodyne has been suggested as a possible technique for the next step in ground-based interferometer arrays at mid-IR wavelengths \cite{monnier_architecture_2016}.
 
In Khabiboulline et al.\cite{khabiboulline_quantum-assisted_2019} (see their Figure 6, panel c) the signal photon, after interaction with the quantum memory, is mixed with an ancillary optical mode (such as from a laser or single-photon source closely matched in frequency) at a beam-splitter, followed by photon counting, to perform the X-basis measurement needed to complete the loading of the signal photon state to the quantum memory, as described in section \ref{sec:qmem_load}. This mixing also serves to hide the ``which-path" information, since the measurement cannot distinguish the signal photon from the ancillary photon(s). If the ancillary mode at each telescope site $A$ and $B$ has a relative phase $\theta_A$ and $\theta_B$, respectively, the resulting state of the quantum memories following loading is
\begin{align}
    \ket{\Psi}
    &= \frac{1}{\sqrt{2}}e^{-i\theta_A}  (\ket{\bar{1
    }_{A} \bar{0}_{B}} \pm e^{i (\phi + (\theta_A- \theta_B))} \ket{\bar{0
    }_{A} \bar{1}_{B}}) 
    \label{eq:qmem_with_phase}
\end{align}
The ancillary photon therefore plays exactly the role of the LO in controlling the phase as in the heterodyne technique. If these phases are stable and known, they can be corrected, as described in the previous section. 

As a note, the similar relation presented earlier in section \ref{sec:qmem_load} in Equation \ref{eq:memory_phase_transfer} suppresses the local loading reference phase for clarity. In that case of a physical reflective-memory implementation, the cavity interaction entangles the stellar mode with the memory, while subsequent interference with a phase-defined ancillary optical mode and measurement of the outgoing light complete the state transfer; the resulting memory-stored inter-site phase includes the difference between the two local reference phases, as shown explicitly in Equation \ref{eq:qmem_with_phase}.

Therefore, quantum-memory systems and heterodyne systems maintain the same requirements for phase precision and control. For example, Bourdarot \cite{bourdarot_heterodyne_2022} envisages LO distribution over optical fibers to provide the phase reference at each site for a km-scale, mid-IR array. In the Stas et al.\cite{Stas_2026} experiment demonstrating interferometry with quantum memories, the laser source stability for both pre-entangling and loading the quantum memories is consistent with our treatment of the phase stability requirement, although not explicitly developed there. The results of this experiment are described in more detail in section \ref{sec:keystone}.

Finally, it is important to note that although both techniques use the LO as phase reference, classical heterodyne and quantum entanglement interferometry remain fundamentally different. As in direct detection, S/N in quantum interferometry scales as $\sqrt{\epsilon}$, whereas coherent detection methods such as heterodyne scale as $\epsilon$, where $\epsilon$ is the photon occupancy number. This critical difference allows quantum interferometry to be viable at visible wavelengths for astronomical sources with $\epsilon \ll \rm1$ where the shot noise would be prohibitive for heterodyne. For example, $\epsilon$ is $\sim \rm{10}^{\rm{-6}}$ for the typical source and aperture parameters adopted in Khabiboulline et al.\cite{khabiboulline_optical_2019, khabiboulline_quantum-assisted_2019} and the best-achievable S/N would be $\sim$ 1000 times better for a quantum system versus heterodyne in that case. For multiple apertures, the quantum technique offers unique advantages, for example the ability to perform joint quantum measurement of the coherence over the array.

\subsection{Requirements}

To conclude this section, we present a table of requirements for a future quantum-memory based long-baseline interferometry system. 

\begin{longtable}{>{\RaggedRight\arraybackslash}p{0.18\textwidth}
                  >{\RaggedRight\arraybackslash}p{0.18\textwidth}
                  >{\RaggedRight\arraybackslash}p{0.22\textwidth}
                  >{\RaggedRight\arraybackslash}p{0.28\textwidth}
                  >{\centering\arraybackslash}p{0.08\textwidth}}
\caption{Critical operational parameters for a quantum-assisted optical interferometer under the fiducial Khabiboulline et al. (2019) assumptions: 10~m$^2$ aperture, $V=10$ mag source, 10~GHz optical bandwidth, central wavelength $\lambda_0 = 633$~nm, coherence length $\ell_c \sim 3$~cm, and 10~km baseline. The values shown are intended as minimal order-of-magnitude system requirements rather than a complete engineering budget. \newline
\textit{*In the table below, $B$ stands for the baseline distance, the physical separation between telescopes.}}
\label{tab:critical_parameters_quantum_interferometer}\\

\toprule
\textbf{Parameter} &
\textbf{Requirement} &
\textbf{How arrived at} &
\textbf{Notes} \\
\midrule
\endfirsthead

\toprule
\textbf{Parameter} &
\textbf{Value} &
\textbf{How arrived at} &
\textbf{Notes} \\
\midrule
\endhead

\bottomrule
\endfoot

Photon occupancy 
(number per inverse bandwidth) &
$10^{-6}$ &
Set by source properties for weak thermal light. &
Photon occupancy per temporal-spectral mode for the assumed source, aperture, and bandwidth.  \\

Photon rate &
$\sim 1~\mathrm{ms}^{-1}$ &
Set by source brightness, aperture, and bandwidth. &
Assumes the fiducial Khabiboulline et al. parameters.  \\

Photon coherence time &
$t_c \sim 10^{-10}~\mathrm{s}$ &
Set by the 10~GHz spectral filter. &
The stellar photon must be well matched to the spectral acceptance profile of the quantum memory; this corresponds to $\ell_c \sim 3$~cm.  \\

Memory Coherence time & $10^{-3}$ s & Set by photon rate

\\

Number of memories per site, for a single wavelength band &
$\sim 20$ &
Set by binary encoding for the assumed photon rate. &
Memory multiplexing supports the required photon-arrival statistics for the single-band case.  \\

Entanglement rate &
$\sim 200~\mathrm{kHz}$ &
Set by photon rate multiplied by the number of memories. &
Required entanglement-distribution rate for the assumed memory multiplexing. \\

LO frequency stability &
$\sim 2\times 10^{4}~\mathrm{Hz}$ at $\nu_0 \simeq 474~\mathrm{THz}$, or $\Delta \nu/\nu \sim 4\times 10^{-11}$ &
Set by angular-resolution tolerance and geometric-delay tracking for a 10~km baseline (B). &
The response varies as $\cos(\omega_0 \Delta t)$; the LO frequency must remain within the tolerance corresponding to $\sim 1.2\lambda/B$.  \\

LO phase stability &
Within $\frac{\lambda}{2\pi}$ over the coherence time &
Set by visibility phase error limits &
Keeps the coherence-envelope response at approximately the 90\% level for the assumed rectangular 10~GHz band.  \\

Timing precision &
$\sim 10^{-11}~\mathrm{s}$ &
Set to approximately one-tenth of the coherence time. &
Consistent with maintaining instrumental visibility at the $\sim 80\%$ level  \\

\end{longtable}

\section{Keystone areas of technology development} \label{sec:keystone}

The keystone technology areas for quantum-memory-assisted nonlocal interferometry are (1) spectrally matched atom-photon interface for astronomically relevant bands, (2) scalable design for a quantum memory hub with long storage time and high-fidelity local qubit control, and (3) phase-stable operation across separated sites and remote apertures. For astronomy, the most relevant figure of merit is the overall probability that an astronomical optical mode is jointly collected, mapped into a countable quantum-memory resource with pre-shared entanglement at remote stations, and read out while preserving the phase information needed for imaging. An applicable testbed therefore requires simultaneous progress in efficiency, bandwidth, coherence, fidelity, phase stability, entanglement rate, and multiplexed scaling.

Recent progress shows that these underlying primitives are feasible in a laboratory environment. Leveraging their developed quantum memory of silicon-vacancy centers in diamond nanocavity, Stas \textit{et al.} demonstrated entanglement-assisted nonlocal optical interferometry over a two-node quantum network with a fiber-link baseline up to 1.55 km \cite{Stas_2026}. Each node contains a quantum memory device, featuring an electronic spin used as a communication qubit and a $^{29}$Si nuclear spin used as a memory qubit. The experiment combines event-ready remote nuclear-spin entanglement and joint memory readout to recover a differential optical phase from weak incident light. They reported the following performance metrics, including a nuclear Bell-state fidelity of $F=0.73(4)$ at 0.25 Hz in the local configuration, a 0.41 Hz entanglement generation rate for the interferometry sequence, an average data-collection rate of approximately 12 mHz with heralding, an interferometric visibility of $0.090(26)$ with nonlocal photon heralding compared with $0.031(18)$ without heralding, and nuclear Bell-state fidelity $F=0.63(3)$. The experiment also reported active phase locking with optical interference visibility around 0.93. These numbers are modest for astronomy, but they are important as a demonstration of the full logical sequence of a memory-assisted nonlocal phase measurement in a solid-state quantum-network architecture.

A complementary experiment demonstrated a memory-assisted nonlocal interferometer using two $^{87}$Rb cold-atomic-ensemble quantum memories \cite{wang2026memory} and DLCZ-type heralded entanglement \cite{duan2001long}. Instead of SiV spin-photon nodes, this implementation uses retrieved atomic-ensemble entanglement photons as the auxiliary nonlocal resource and measured a simulated thermal light field generated by Raman scattering. It achieved an equivalent fiber-link baseline of 20 km using quantum frequency conversion of write-out photons from 780 nm to 1522 nm, while also demonstrating compensation of a geometric delay by delaying memory readout by 5 $\mu$s. Reported metrics include retrieval efficiency $\eta_{\rm ro}\simeq 26\%$, coherence time $15.4\pm0.3$ ns, local visibility $0.54\pm0.05$, visibility over 20 km-baseline $0.51\pm0.04$ with a 20 ns window, $0.39\pm0.01$ with a 60 ns window, and $0.32\pm0.03$ when simulating the delayed-arrival case. This experiment represents a step forward toward astronomy applications as it uses a thermal-light model and addresses a longer fiber baseline with geometric-delay compensation.

Together, these two demonstrations define the present frontier. The SiV-in-diamond experiment is closest to the solid-state quantum-network vision: the entangled memory pair is prepared before the sensing attempt, the incident optical mode is locally processed and erased without revealing which aperture received the photon, and the resulting phase information is stored in matter qubits for later readout. In this event-ready architecture, the critical requirements are: (i) a bright, coherent, and spectrally stable spin-photon interface, usually enhanced by nanophotonic cavities or high-collection optics, to generate spin--photon entanglement and efficient photon-memory gates \cite{togan2010quantum,bernien2013heralded,nguyen2019quantum,knaut2024entanglement}; (ii) a long-lived multi-qubit memory register\cite{bradley2019ten} with high-fidelity local gates and single-shot readout, so that heralded entanglement can survive feed-forward, local processing, and delayed readout \cite{robledo2011high,pompili2021realization,stas2022robust}; and (iii) a low-loss telecom-compatible network layer with indistinguishable photons, active phase stabilization or phase-insensitive time-bin protocols, real-time heralding, and multiplexing to raise the event-ready entanglement rate \cite{tchebotareva2019entanglement,stolk2024metropolitan,knaut2024entanglement,ruskuc2025multiplexed}. Progress across NV centers in diamond\cite{togan2010quantum,bernien2013heralded,pompili2021realization}, SiV centers in SiC\cite{nagy2019high,nagy2018quantum,fang2024experimental}, and rare-earth-ion memories shows that these primitives can be implemented in complementary solid-state platforms \cite{lago2021telecom,liu2021heralded,tittel2025quantum}. The cold-ensemble-based DLCZ experiment is closer to the astronomical-interferometry use case in two other engineering aspects: it uses a thermal-light model and reaches a 20 km equivalent fiber baseline with explicit geometric-delay compensation. Taken together, these demonstrations show that the core physics of memory-assisted nonlocal interferometry is feasible, but astronomy-grade operation still requires much higher throughput, broader or multiplexed spectral acceptance, lower-noise frequency conversion, stronger phase stability, and scalable multi-aperture entanglement distribution \cite{Stas_2026,wang2026memory}. 

The remaining gap is primarily caused by performance and scalability of the quantum memory. Astronomical sources are weak and broadband, so optical loss and narrow acceptance bandwidth in both discussed demonstrations are bottlenecks. For either approach to become an astronomical instrument, the system must support much higher remote entanglement generation throughput\cite{Cui2025CSA,DharaCapacity2025,Winnel2022}, broader or heavily multiplexed spectral operation\cite{anand2026programmable}, low-noise high-efficiency frequency conversion where needed\cite{bersin2024telecom}, and better phase stabilization with a full field-test error budget\cite{wang2026memory}.

Near-term pathfinders should therefore focus on system-level improvement for loading real stellar light into a memory-assisted nonlocal interferometer and quantify the complete efficiency, bandwidth, phase, and operation time budget. The next imminent milestone demonstration would be to test such an upgraded system under realistic situations, including low photon fluxes with broader optical bandwidths, and a variety of environmental noise sources. Full astronomical deployment remains a longer-term goal requiring multiplexed memory registers, high-rate event-ready entanglement generation across multi-apertures, and sustained quantum-network development.
\section{Conclusions and Outlook} \label{sec:conclusions}

\subsection{Conclusions}

Classical optical interferometry has provided some of the highest resolution astronomical observations to date. However, pushing towards higher resolutions suffers from considerable engineering challenges in the classical case, most notably the necessity of both transporting the photon wavepackets to the same location to be interfered and the synchronization of these wavepackets. In this paper, we have confirmed that there is an alternate path towards achieving interferometry at baselines far beyond the capacity of classical interferometry by leveraging quantum memory.

Quantum-memory interferometry changes the physical location and form of the interferometric measurement.  The incident optical coherence is mapped locally into phase-referenced quantum memories and subsequently interrogated using entanglement-assisted operations. In this sense, quantum memory replaces physical photon transport and full-range optical delay lines with memory loading, time-bin association, quantum gates, and delayed readout.

The central conclusion of this paper is that the relevant stability requirements are transformed in the quantum realization, offering a potential path towards significantly longer baselines. In a classical direct-detection interferometer, the same optical path must both overlap the coherence envelopes and maintain sub-wavelength phase stability during beam combination. In a quantum-memory architecture, these requirements are separated. The timing problem is primarily the problem of capturing the same stellar coherence packet into corresponding memory modes at different sites, while the phase problem is primarily the problem of maintaining and calibrating the local phase references that define the memory-loading frame.

The two main mathematical findings can be summarized as follows:

\begin{itemize}
    \item \textbf{Timing requirements in the quantum realization are set by coherence-envelope overlap, time-bin association, and memory operation speed.}
    For a filtered stellar photon with coherence time $t_c$ loaded into memory exposure windows of duration $t_w$, the capture efficiency and recovered visibility are controlled by the overlap between the photon wave packet and the temporally gated memory mode. The relevant synchronization parameter is the residual time-bin error $\tau$, and the visibility loss is governed by the normalized envelope-overlap factor $f(\tau)/f(0)$. Thus the timing requirement is not sub-wavelength path control at the optical carrier scale $\lambda_0/c$ as in the classical case, but rather accurate association of corresponding temporal modes on the scale set by $t_c$, $t_w$, and the memory gate/readout operations, reducing the required precision by orders of magnitude for this aspect of the measurement.

    \item \textbf{Phase requirements are set by local phase-reference stability, geometric-delay correction, and ordinary interferometric nuisance terms.}
    The recovered complex visibility retains the optical phase information through the memory-loading reference frame. The measured phase includes the astronomical visibility phase together with deterministic geometric-delay terms, local oscillator phase, and atmospheric/instrumental phase errors. Known geometric contributions can in principle be corrected through phase operations on the memories, while stochastic nuisance terms such as atmospheric piston must still be measured, calibrated, or removed through interferometric observables such as closure phase. Quantum memory therefore relaxes the need for kilometer-scale optical delay control, but it does not remove the need for phase referencing, phase calibration, or atmospheric-error mitigation that are present in the classical case as well.
\end{itemize}

This separation of timing and phase requirements is closely analogous to the distinction between direct-detection and heterodyne interferometry. In heterodyne interferometry, the optical field is mixed locally with a phase-stable local oscillator, and the interferometric phase is carried in a lower-frequency reference frame before electronic correlation. In the quantum-memory case, the ancillary optical mode used during memory loading plays a similar role: it defines the local phase reference with respect to which the stellar coherence is written into matter degrees of freedom. The subsequent measurement is not a classical heterodyne measurement, and it preserves the favorable photon-counting scaling of quantum interferometry in the weak-light regime, but the analogy is useful for understanding where the phase-reference requirement enters.

Quantum-memory interferometry therefore changes the engineering requirements for long-baseline optical interferometry rather than the underlying interferometric problem. It can remove the need to physically transport astronomical photons, substantially reduce or eliminate full-range optical delay lines, and open the possibility of memory-based processing of the captured optical coherence. At the same time, it preserves the familiar requirements of visibility reconstruction: coherence must be captured, phase must be referenced, deterministic delay must be corrected, and atmospheric and instrumental errors must be calibrated. If the required memory bandwidth, storage time, loading efficiency, phase stability, and entanglement distribution rates can be achieved, these architectures provide a plausible route toward optical interferometric baselines far beyond those accessible with classical beam transport alone.

\subsection{Outlook}

Quantum information technologies reduce or eliminate some of the most difficult hurdles in the way of creating very long-baseline optical interferometers.  However, significant technical barriers remain in the development of quantum memories and quantum repeaters before they are ready for use in a high-performance interferometric array.  In this paper, we have discussed the performance requirements for quantum memories, which should help direct research progress towards better quantum networking devices.

While quantum technology can address some of the challenges of very long-baseline optical interferometers, some challenges are shared by classical and quantum techniques.  The quantum approach we have discussed requires the sites in the array to share phase-locked local oscillators, and synchronizing the local oscillators between sites is not straightforward, although not as difficult as preserving coherence of single photons over long distances.  Atmospheric fluctuations pose an equal challenge for classical and quantum measurements of the complex visibility between sites.  The methods used to address atmospheric fluctuation in classical interferometers, such as phase closure, can also be applied to interferometers based on quantum memories\cite{Gorshkov2026_SPIE}.
Quantum information approaches may offer other advantages as well --- once photons are captured in a quantum memory, a much wider array of quantum algorithms could in principle be applied, potentially allowing measurements that would be inaccessible to a more classical device. Near-term opportunities exist in hybrid systems. Long-term potential for fully quantum interferometric arrays over large distances will only follow significant developments in quantum technologies.

\acknowledgments  

J.G.R. was supported by a NASA Space Technology Graduate Research Opportunity. We gratefully acknowledge support from the NSF Engineering Research Center for Quantum Networks (Grant No. EEC-1941583). W.D. was supported by NSF grant PHY-2210361 and the Maryland Center for Fundamental Physics. D.G. and Z.L. acknowledge support from the National Science Foundation (QLCI grant OMA-2120757). The work of B.M.M., J.R., and S.T.R. is supported by NOIRLab, which is managed by the
Association of Universities for Research in Astronomy (AURA) under a cooperative agreement with the U.S. National Science Foundation. E.T.K. was supported in part by ONR MURI (award No. N000142612102). 

We acknowledge the use of AI tools in the preparation of this manuscript. Specifically, ChatGPT 5.5 and Claude Opus 4.8 were used to improve the wording and clarity of selected passages originally written by the authors and to assist certain authors in understanding technical concepts. All AI-assisted revisions were carefully reviewed, verified, and further edited by the authors to ensure the accuracy and appropriateness of the final text.

\bibliography{report} 

@article{Gottesman2012LongBaseline,
  title = {Longer-Baseline Telescopes Using Quantum Repeaters},
  author = {Gottesman, Daniel and Jennewein, Thomas and Croke, Sarah},
  journal = {Phys. Rev. Lett.},
  volume = {109},
  issue = {7},
  pages = {070503},
  numpages = {5},
  year = {2012},
  month = {Aug},
  publisher = {American Physical Society},
  doi = {10.1103/PhysRevLett.109.070503},
  url = {https://link.aps.org/doi/10.1103/PhysRevLett.109.070503}
}

@article{khabiboulline_quantum-assisted_2019,
    title = {Quantum-assisted telescope arrays},
    volume = {100},
    issn = {2469-9926, 2469-9934},
    url = {https://link.aps.org/doi/10.1103/PhysRevA.100.022316},
    doi = {10.1103/PhysRevA.100.022316},
    language = {en},
    number = {2},
    urldate = {2024-09-05},
    journal = {Physical Review A},
    author = {Khabiboulline, E. T. and Borregaard, J. and De Greve, K. and Lukin, M. D.},
    month = aug,
    year = {2019},
    pages = {022316},
}

@article{khabiboulline_optical_2019,
    title = {Optical {Interferometry} with {Quantum} {Networks}},
    volume = {123},
    issn = {0031-9007, 1079-7114},
    url = {https://link.aps.org/doi/10.1103/PhysRevLett.123.070504},
    doi = {10.1103/PhysRevLett.123.070504},
    language = {en},
    number = {7},
    urldate = {2024-09-06},
    journal = {Physical Review Letters},
    author = {Khabiboulline, E.T. and Borregaard, J. and De Greve, K. and Lukin, M.D.},
    month = aug,
    year = {2019},
    pages = {070504},
}

@book{NielsenChuang2010, 
    place={Cambridge}, 
    title={Quantum Computation and Quantum Information: 10th Anniversary Edition},
    publisher={Cambridge University Press}, 
    author={Nielsen, Michael A. and Chuang, Isaac L.}, 
    year={2010}
}

@article{DuanKimble2004,
  title = {Scalable Photonic Quantum Computation through Cavity-Assisted Interactions},
  author = {Duan, L.-M. and Kimble, H. J.},
  journal = {Phys. Rev. Lett.},
  volume = {92},
  issue = {12},
  pages = {127902},
  numpages = {4},
  year = {2004},
  month = {Mar},
  publisher = {American Physical Society},
  doi = {10.1103/PhysRevLett.92.127902},
  url = {https://link.aps.org/doi/10.1103/PhysRevLett.92.127902}
}

@article{Sajjad2026-single-rail,
      title={Boosted linear-optical measurements on single-rail qubits with unentangled ancillas}, 
      author={Aqil Sajjad and Isack Padilla and Saikat Guha},
      year={2026},
      eprint={2603.16795},
      archivePrefix={arXiv},
      primaryClass={quant-ph},
      url={https://arxiv.org/abs/2603.16795}, 
}

@BOOK{Kok2010,
  title     = "Introduction to Optical Quantum Information Processing",
  author    = "Kok, Pieter and Lovett, Brendon W",
  publisher = "Cambridge University Press",
  month     =  apr,
  year      =  2010
}

@article{GRISSANCHEZ2016,
author = {I. Gris-S\'{a}nchez and D. Van Ras and T. A. Birks},
journal = {Optica},
number = {3},
pages = {270--276},
publisher = {OSA},
title = {The Airy fiber: an optical fiber that guides light diffracted by a circular aperture},
volume = {3},
month = {Mar},
year = {2016},
url = {http://www.osapublishing.org/optica/abstract.cfm?URI=optica-3-3-270},
doi = {10.1364/OPTICA.3.000270},
}

@article{duan2001long,
  title={Long-distance quantum communication with atomic ensembles and linear optics},
  author={Duan, L-M and Lukin, Mikhail D and Cirac, J Ignacio and Zoller, Peter},
  journal={Nature},
  volume={414},
  number={6862},
  pages={413--418},
  year={2001},
  publisher={Nature Publishing Group UK London}
}

@article{pompili2021realization,
  title={Realization of a multinode quantum network of remote solid-state qubits},
  author={Pompili, Matteo and Hermans, Sophie LN and Baier, Simon and Beukers, Hans KC and Humphreys, Peter C and Schouten, Raymond N and Vermeulen, Raymond FL and Tiggelman, Marijn J and dos Santos Martins, Laura and Dirkse, Bas and others},
  journal={Science},
  volume={372},
  number={6539},
  pages={259--264},
  year={2021},
  publisher={American Association for the Advancement of Science}
}

@article{bernien2013heralded,
  title={Heralded entanglement between solid-state qubits separated by three metres},
  author={Bernien, Hannes and Hensen, Bas and Pfaff, Wolfgang and Koolstra, Gerwin and Blok, Machiel S and Robledo, Lucio and Taminiau, Tim H and Markham, Matthew and Twitchen, Daniel J and Childress, Lilian and others},
  journal={Nature},
  volume={497},
  number={7447},
  pages={86--90},
  year={2013},
  publisher={Nature Publishing Group UK London}
}

@article{nagy2019high,
  title={High-fidelity spin and optical control of single silicon-vacancy centres in silicon carbide},
  author={Nagy, Roland and Niethammer, Matthias and Widmann, Matthias and Chen, Yu-Chen and Udvarhelyi, P{\'e}ter and Bonato, Cristian and Hassan, Jawad Ul and Karhu, Robin and Ivanov, Ivan G and Son, Nguyen Tien and others},
  journal={Nature communications},
  volume={10},
  number={1},
  pages={1954},
  year={2019},
  publisher={Nature Publishing Group UK London}
}

@article{nagy2018quantum,
  title={Quantum properties of dichroic silicon vacancies in silicon carbide},
  author={Nagy, Roland and Widmann, Matthias and Niethammer, Matthias and Dasari, Durga BR and Gerhardt, Ilja and Soykal, {\"O}ney O and Radulaski, Marina and Ohshima, Takeshi and Vu{\v{c}}kovi{\'c}, Jelena and Son, Nguyen Tien and others},
  journal={Physical Review Applied},
  volume={9},
  number={3},
  pages={034022},
  year={2018},
  publisher={APS}
}

@article{tittel2025quantum,
  title={Quantum networks using rare-earth ions},
  author={Tittel, Wolfgang and Afzelius, Mikael and Kinos, Adam and Rippe, Lars and Walther, Andreas},
  journal={Quantum Science and Technology},
  volume={10},
  number={3},
  pages={033002},
  year={2025},
  publisher={IOP Publishing}
}

@article{robledo2011high,
  title={High-fidelity projective read-out of a solid-state spin quantum register},
  author={Robledo, Lucio and Childress, Lilian and Bernien, Hannes and Hensen, Bas and Alkemade, Paul FA and Hanson, Ronald},
  journal={Nature},
  volume={477},
  number={7366},
  pages={574--578},
  year={2011},
  publisher={Nature Publishing Group UK London}
}

@article{togan2010quantum,
  title={Quantum entanglement between an optical photon and a solid-state spin qubit},
  author={Togan, Emre and Chu, Yiwen and Trifonov, Alexei S and Jiang, Liang and Maze, Jeronimo and Childress, Lilian and Dutt, MV Gurudev and S{\o}rensen, Anders S{\o}ndberg and Hemmer, Phillip R and Zibrov, Alexander S and others},
  journal={Nature},
  volume={466},
  number={7307},
  pages={730--734},
  year={2010},
  publisher={Nature Publishing Group UK London}
}

@article{nguyen2019quantum,
  title={Quantum network nodes based on diamond qubits with an efficient nanophotonic interface},
  author={Nguyen, CT and Sukachev, DD and Bhaskar, MK and Machielse, Bartholomeus and Levonian, DS and Knall, EN and Stroganov, Pavel and Riedinger, Ralf and Park, Hongkun and Lon{\v{c}}ar, M and others},
  journal={Physical review letters},
  volume={123},
  number={18},
  pages={183602},
  year={2019},
  publisher={APS}
}

@article{knaut2024entanglement,
  title={Entanglement of nanophotonic quantum memory nodes in a telecom network},
  author={Knaut, Can M and Suleymanzade, Aziza and Wei, Y-C and Assumpcao, Daniel R and Stas, P-J and Huan, Yan Qi and Machielse, Bartholomeus and Knall, Erik N and Sutula, Madison and Baranes, Gefen and others},
  journal={Nature},
  volume={629},
  number={8012},
  pages={573--578},
  year={2024},
  publisher={Nature Publishing Group UK London}
}

@article{stas2022robust,
  title={Robust multi-qubit quantum network node with integrated error detection},
  author={Stas, P-J and Huan, Yan Qi and Machielse, Bartholomeus and Knall, Erik N and Suleymanzade, Aziza and Pingault, Benjamin and Sutula, Madison and Ding, Sophie W and Knaut, Can M and Assumpcao, Daniel R and others},
  journal={Science},
  volume={378},
  number={6619},
  pages={557--560},
  year={2022},
  publisher={American Association for the Advancement of Science}
}

@article{bradley2019ten,
  title={A ten-qubit solid-state spin register with quantum memory up to one minute},
  author={Bradley, Conor E and Randall, Joe and Abobeih, Mohamed H and Berrevoets, Remon C and Degen, Maarten J and Bakker, Michiel A and Markham, Matthew and Twitchen, Daniel J and Taminiau, Tim H},
  journal={Physical Review X},
  volume={9},
  number={3},
  pages={031045},
  year={2019},
  publisher={APS}
}

@article{tchebotareva2019entanglement,
  title={Entanglement between a diamond spin qubit and a photonic time-bin qubit at telecom wavelength},
  author={Tchebotareva, Anna and Hermans, Sophie LN and Humphreys, Peter C and Voigt, Dirk and Harmsma, Peter J and Cheng, Lun K and Verlaan, Ad L and Dijkhuizen, Niels and De Jong, Wim and Dr{\'e}au, Ana{\"\i}s and others},
  journal={arXiv preprint arXiv:1905.08676},
  year={2019}
}

@article{stolk2024metropolitan,
  title={Metropolitan-scale heralded entanglement of solid-state qubits},
  author={Stolk, Arian J and van der Enden, Kian L and Slater, Marie-Christine and te Raa-Derckx, Ingmar and Botma, Pieter and Van Rantwijk, Joris and Biemond, JJ Benjamin and Hagen, Ronald AJ and Herfst, Rodolf W and Koek, Wouter D and others},
  journal={Science advances},
  volume={10},
  number={44},
  pages={eadp6442},
  year={2024},
  publisher={American Association for the Advancement of Science}
}

@article{ruskuc2025multiplexed,
  title={Multiplexed entanglement of multi-emitter quantum network nodes},
  author={Ruskuc, Andrei and Wu, C-J and Green, Emanuel and Hermans, Sophie LN and Pajak, William and Choi, Joonhee and Faraon, Andrei},
  journal={Nature},
  volume={639},
  number={8053},
  pages={54--59},
  year={2025},
  publisher={Nature Publishing Group UK London}
}

@article{fang2024experimental,
  title={Experimental generation of spin-photon entanglement in silicon carbide},
  author={Fang, Ren-Zhou and Lai, Xiao-Yi and Li, Tao and Su, Ren-Zhu and Lu, Bo-Wei and Yang, Chao-Wei and Liu, Run-Ze and Qiao, Yu-Kun and Li, Cheng and He, Zhi-Gang and others},
  journal={Physical Review Letters},
  volume={132},
  number={16},
  pages={160801},
  year={2024},
  publisher={APS}
}

@article{lago2021telecom,
  title={Telecom-heralded entanglement between multimode solid-state quantum memories},
  author={Lago-Rivera, Dario and Grandi, Samuele and Rakonjac, Jelena V and Seri, Alessandro and De Riedmatten, Hugues},
  journal={Nature},
  volume={594},
  number={7861},
  pages={37--40},
  year={2021},
  publisher={Nature Publishing Group UK London}
}

@article{liu2021heralded,
  title={Heralded entanglement distribution between two absorptive quantum memories},
  author={Liu, Xiao and Hu, Jun and Li, Zong-Feng and Li, Xue and Li, Pei-Yun and Liang, Peng-Jun and Zhou, Zong-Quan and Li, Chuan-Feng and Guo, Guang-Can},
  journal={Nature},
  volume={594},
  number={7861},
  pages={41--45},
  year={2021},
  publisher={Nature Publishing Group UK London}
}

@inproceedings{bourdarot_heterodyne_2022,
    title = {Heterodyne interferometry: review and prospects},
    volume = {12183},
    shorttitle = {Heterodyne interferometry},
    url = {https://www.spiedigitallibrary.org/conference-proceedings-of-spie/12183/1218312/Heterodyne-interferometry-review-and-prospects/10.1117/12.2635601},
    doi = {10.1117/12.2635601},
    language = {en},
    urldate = {2026-06-23},
    booktitle = {Optical and {Infrared} {Interferometry} and {Imaging} {VIII}},
    publisher = {SPIE},
    author = {Bourdarot, Guillaume},
    month = aug,
    year = {2022},
    pages = {390},
}

@inproceedings{monnier_architecture_2016,
    title = {Architecture design study and technology road map for the {Planet} {Formation} {Imager} ({PFI})},
    volume = {9907},
    url = {https://www.spiedigitallibrary.org/conference-proceedings-of-spie/9907/99071O/Architecture-design-study-and-technology-road-map-for-the-Planet/10.1117/12.2233311},
    doi = {10.1117/12.2233311},
    language = {en},
    urldate = {2026-06-23},
    booktitle = {Optical and {Infrared} {Interferometry} and {Imaging} {V}},
    publisher = {SPIE},
    author = {Monnier, John D. and Ireland, Michael J. and Kraus, Stefan and Baron, Fabien and Creech-Eakman, Michelle and Dong, Ruobing and Isella, Andrea and Merand, Antoine and Michael, Ernest and Minardi, Stefano and Mozurkewich, David and Petrov, Romain and Rinehart, Stephen and Brummelaar, Theo ten and Vasisht, Gautam and Wishnow, Ed and Young, John and Zhu, Zhaohuan},
    month = aug,
    year = {2016},
    pages = {478},
}

@misc{Cui2025CSA,
  author={Cui, Chaohan and Dhara, Prajit and Guha, Saikat},
  title={Coherent State Assisted Entanglement Generation Between Quantum Memories},
  eprint={2504.20344},
  archivePrefix={arXiv},
  primaryClass={quant-ph},
  year={2025}
}

@article{DharaCapacity2025,
  author={Dhara, Prajit and Jiang, Liang and Guha, Saikat},
  title={Entangling quantum memories at channel capacity},
  journal={Optica Quantum},
  volume={3},
  pages={480--486},
  year={2025},
  doi={10.1364/OPTICAQ.570931}
}

@article{Winnel2022,
  author={Winnel, M. S. and Guanzon, J. J. and Hosseinidehaj, N. and Ralph, T. C.},
  title={Achieving the ultimate end-to-end rates of lossy quantum communication networks},
  journal={npj Quantum Information},
  volume={8},
  pages={129},
  year={2022}
}

@article{anand2026programmable,
  title={Programmable Quantum Matter: Heralding Large Cluster States in Driven Inhomogeneous Spin Ensembles},
  author={Anand, Pratyush and Follet, Louis and Hooybergs, Odiel and Englund, Dirk R},
  journal={PRX Quantum},
  volume={7},
  number={2},
  pages={020362},
  year={2026},
  publisher={APS}
}

@article{bersin2024telecom,
  title={Telecom networking with a diamond quantum memory},
  author={Bersin, Eric and Sutula, Madison and Huan, Yan Qi and Suleymanzade, Aziza and Assumpcao, Daniel R and Wei, Yan-Cheng and Stas, Pieter-Jan and Knaut, Can M and Knall, Erik N and Langrock, Carsten and others},
  journal={PRX Quantum},
  volume={5},
  number={1},
  pages={010303},
  year={2024},
  publisher={APS}
}

@article{Bland-Hawthorn_2021,
author = {Joss Bland-Hawthorn and Matthew J. Sellars and John G. Bartholomew},
journal = {J. Opt. Soc. Am. B},
number = {7},
pages = {A86--A98},
publisher = {Optica Publishing Group},
title = {Quantum memories and the double-slit experiment: implications for astronomical interferometry},
volume = {38},
month = {Jul},
year = {2021},
url = {https://opg.optica.org/josab/abstract.cfm?URI=josab-38-7-A86},
doi = {10.1364/JOSAB.424651},
}

@article{Brown_2023,
  title = {Interferometric Imaging Using Shared Quantum Entanglement},
  author = {Brown, Matthew R. and Allgaier, Markus and Thiel, Val\'erian and Monnier, John D. and Raymer, Michael G. and Smith, Brian J.},
  journal = {Phys. Rev. Lett.},
  volume = {131},
  issue = {21},
  pages = {210801},
  numpages = {6},
  year = {2023},
  month = {Nov},
  publisher = {American Physical Society},
  doi = {10.1103/PhysRevLett.131.210801},
  url = {https://link.aps.org/doi/10.1103/PhysRevLett.131.210801}
}

@article{Stas_2026,
  title = {Entanglement-assisted non-local optical interferometry in a quantum network},
  author = {Stas, P.-J. and Wei, Y.-C. and Sirotin, M. and Huan, Y. Q. and Yazlar, U. and Abdo Arias, F. and Knyazev, E. and Baranes, G. and Machielse, B. and Grandi, S. and Riedel, D. and Borregaard, J. and Park, H. and Lončar, M. and Suleymanzade, A. and Lukin, M. D.},
  year = {2026},
  journal = {Nature},
  volume = {651},
  number = {8105},
  pages = {326--332},
  doi = {10.1038/s41586-026-10171-w},
  url = {https://doi.org/10.1038/s41586-026-10171-w},
  isbn = {1476-4687}
}

@article{wang2026memory,
  title={Memory-Assisted Nonlocal Interferometer toward Long-Baseline Telescopes},
  author={Wang, Bin and Luo, Xi-Yu and Gao, Bo-Feng and Liu, Jian-Long and Wang, Chao-Yang and Yan, Zi and Ke, Qiao-Mu and Teng, Da and Zheng, Ming-Yang and Cao, Yuan and others},
  journal={Physical Review Letters},
  volume={136},
  number={24},
  pages={240801},
  year={2026},
  publisher={APS}

}

@book{Goodman_2005,
  title={Introduction to Fourier Optics},
  author={Goodman, J.W.},
  isbn={9780974707723},
  lccn={2004023213},
  series={McGraw-Hill physical and quantum electronics series},
  url={https://books.google.com/books?id=ow5xs_Rtt9AC},
  year={2005},
  publisher={W. H. Freeman}
}

@misc{Liu2026TimeDelays,
      title={Measuring gravitational lensing time delays with quantum information processing}, 
      author={Zhenning Liu and William DeRocco and Shiming Gu and Emil T. Khabiboulline and Soonwon Choi and Andrew M. Childs and Anson Hook and Alexey V. Gorshkov and Daniel Gottesman},
      year={2026},
      eprint={2510.07898},
      archivePrefix={arXiv},
      primaryClass={quant-ph},
      url={https://arxiv.org/abs/2510.07898}, 
}

@article{knill_scheme_2001,
    title = {A scheme for efficient quantum computation with linear optics},
    volume = {409},
    copyright = {2001 Macmillan Magazines Ltd.},
    issn = {1476-4687},
    url = {https://www.nature.com/articles/35051009},
    doi = {10.1038/35051009},
    language = {en},
    number = {6816},
    urldate = {2026-06-25},
    journal = {Nature},
    publisher = {Nature Publishing Group},
    author = {Knill, E. and Laflamme, R. and Milburn, G. J.},
    month = jan,
    year = {2001},
    pages = {46--52},
}

@article{Sajjad2024,
 title = {Quantum limits of parameter estimation in long-baseline imaging},
  author = {Sajjad, Aqil and Grace, Michael R. and Guha, Saikat},
  journal = {Phys. Rev. Res.},
  volume = {6},
  issue = {1},
  pages = {013212},
  numpages = {25},
  year = {2024},
  month = {Feb},
  publisher = {American Physical Society},
  doi = {10.1103/PhysRevResearch.6.013212},
  url = {https://link.aps.org/doi/10.1103/PhysRevResearch.6.013212}
}

@article{Padilla2024,
  author        = {Padilla, Isack and Sajjad, Aqil and Saif, Babak N. and Guha, Saikat},
  title         = {{Quantum resolution limit of long-baseline imaging using distributed entanglement}},
  eprint        = {2406.16789},
  archivePrefix = {arXiv},
  primaryClass  = {quant-ph},
  doi           = {10.1103/4npc-gpvh},
  journal       = {Phys. Rev. A},
  volume        = {113},
  number        = {1},
  pages         = {012608},
  year          = {2026}
}

@article{Padilla2025,
  author        = {Padilla, Isack and Sajjad, Aqil and Saif, Babak N. and Guha, Saikat},
  title         = {{Superresolution Imaging with Entanglement-Enhanced Telescopy}},
  eprint        = {2504.03117},
  archivePrefix = {arXiv},
  primaryClass  = {quant-ph},
  doi           = {10.1103/354q-ch63},
  journal       = {Phys. Rev. Lett.},
  volume        = {136},
  number        = {1},
  pages         = {010803},
  year          = {2026}
}

@Book{Helstrom1976,
  Title                    = {Quantum Detection and Estimation Theory},
  Author                   = {Carl. W. Helstrom},
  Publisher                = {Academic Press, New York},
  Year                     = {1976}
}

@inproceedings{Gorshkov2026_SPIE,
  author = {Gorshkov, Alexey V. and
            Gottesman, Daniel and
            Khabiboulline, Emil T. and
            Liu, Zhenning and
            McClinton, Brittany and
            Rajagopal, Jayadev and
            Ridgway, Stephen T.},
  title = {Quantum Closure Phase for Robust Imaging at the Single-Photon Level},
  booktitle = {Optical and Infrared Interferometry and Imaging X},
  series = {Proceedings of SPIE},
  address = {Copenhagen, Denmark},
  month = jul,
  year = {2026},
  note = {To appear}
}

@ARTICLE{Personick1971,
  author={Personick, S.},
  journal={IEEE Transactions on Information Theory}, 
  title={Application of quantum estimation theory to analog communication over quantum channels}, 
  year={1971},
  volume={17},
  number={3},
  pages={240-246},
  doi={10.1109/TIT.1971.1054643}
}

@article{Deshler-unpublished,
  author = {Deshler, Nico and Grace, Michael R. and Ashok, Amit and Guha, Saikat},
  title  = {Quantum Limits of Brightness Estimation for Incoherent Imaging},
  note   = {Unpublished manuscript}
}
\bibliographystyle{spiebib} 

\end{document}